\documentclass[12pt]{article}

\begin{document}

\author{G.Quznetsov \\
quznets@geocities.com}
\title{Probability and Dirac Equation}
\maketitle

\begin{abstract}
The Dirac equation is deduced from the probability properties by the
Poincare group transformations.
\end{abstract}

I'm use notation:

\[
\overrightarrow{x}=\left( x,y,z\right) \mbox{,} 
\]

\[
\beta _x=\left[ 
\begin{array}{cccc}
0 & 1 & 0 & 0 \\ 
1 & 0 & 0 & 0 \\ 
0 & 0 & 0 & -1 \\ 
0 & 0 & -1 & 0
\end{array}
\right] \mbox{, }\beta _y=\left[ 
\begin{array}{cccc}
0 & -i & 0 & 0 \\ 
i & 0 & 0 & 0 \\ 
0 & 0 & 0 & i \\ 
0 & 0 & -i & 0
\end{array}
\right] \mbox{,} 
\]

\[
\beta _z=\left[ 
\begin{array}{cccc}
1 & 0 & 0 & 0 \\ 
0 & -1 & 0 & 0 \\ 
0 & 0 & -1 & 0 \\ 
0 & 0 & 0 & 1
\end{array}
\right] \mbox{.} 
\]

\[
\gamma ^0=\left[ 
\begin{array}{cccc}
0 & 0 & 1 & 0 \\ 
0 & 0 & 0 & 1 \\ 
1 & 0 & 0 & 0 \\ 
0 & 1 & 0 & 0
\end{array}
\right] \mbox{ and }\beta ^4=\left[ 
\begin{array}{cccc}
0 & 0 & i & 0 \\ 
0 & 0 & 0 & i \\ 
-i & 0 & 0 & 0 \\ 
0 & -i & 0 & 0
\end{array}
\right] \mbox{.} 
\]

For every trackelike \cite{L1} probability density $3+1$-vector

\[
\left\langle \rho \left( t,\overrightarrow{x}\right) ,j_x\left( t,%
\overrightarrow{x}\right) ,j_y\left( t,\overrightarrow{x}\right) ,j_z\left(
t,\overrightarrow{x}\right) \right\rangle 
\]

the 4-components complex vector function (4-spinor) $\Psi \left( t,%
\overrightarrow{x}\right) $ exists, for which:

\begin{equation}
\begin{array}{c}
\rho =\Psi ^{\dagger }\cdot \Psi \mbox{,} \\ 
j_x=\Psi ^{\dagger }\cdot \beta _x\cdot \Psi \mbox{, }j_y=\Psi ^{\dagger
}\cdot \beta _y\cdot \Psi \mbox{, }j_z=\Psi ^{\dagger }\cdot \beta _z\cdot
\Psi \mbox{,}
\end{array}
\label{pp}
\end{equation}

The operator $\widehat{U}\left( \triangle t\right) $, which acts in the set
of these spinors, is denoted as the evolution operator for the spinor $\Psi
\left( t,\overrightarrow{x}\right) $, if:

\[
\Psi \left( t+\triangle t,\overrightarrow{x}\right) =\widehat{U}\left(
\triangle t\right) \Psi \left( t,\overrightarrow{x}\right) \mbox{.} 
\]

$\widehat{U}\left( \triangle t\right) $ is a linear operator.

The set of the spinors, for which $\widehat{U}\left( \triangle t\right) $ is
the evolution operator, is denoted as the operator $\widehat{U}\left(
\triangle t\right) $ space.

The operator space is the linear space.

Let for an infinitesimal $\triangle t$:

\[
\widehat{U}\left( \triangle t\right) =1+\triangle t\cdot i\cdot \widehat{H}%
\mbox{.} 
\]

Hence for an elements of the operator $\widehat{U}\left( \triangle t\right) $
space:

\[
i\cdot \widehat{H}=\partial _t\mbox{.} 
\]

Since the functions $\rho $, $j_x$, $j_y$, $j_z$ fulfill to the continuity
equation:

\[
\partial _t\rho +\partial _xj_x+\partial _yj_y+\partial _zj_z=0
\]

then:

\[
\left( \left( \partial _t\Psi ^{\dagger }\right) +\left( \partial _x\Psi
^{\dagger }\right) \cdot \beta _x+\left( \partial _y\Psi ^{\dagger }\right)
\cdot \beta _y+\left( \partial _z\Psi ^{\dagger }\right) \cdot \beta
_z\right) \cdot \Psi = 
\]

\[
=-\Psi ^{\dagger }\cdot \left( \left( \partial _t+\beta _x\cdot \partial
_x+\beta _y\cdot \partial _y+\beta _z\cdot \partial _z\right) \Psi \right) %
\mbox{.} 
\]

Let:

\[
\widehat{Q}=\left( i\cdot \widehat{H}+\beta _x\cdot \partial _x+\beta
_y\cdot \partial _y+\beta _z\cdot \partial _z\right) \mbox{.} 
\]

Hence:

\[
\Psi ^{\dagger }\cdot \widehat{Q}^{\dagger }\cdot \Psi =-\Psi ^{\dagger
}\cdot \widehat{Q}\cdot \Psi \mbox{.} 
\]

Therefore $i\cdot \widehat{Q}$ is the Hermitean operator.

Hence $\widehat{H}$ is the Hermitean operator, too. And $\widehat{U}\left(
\triangle t\right) $ is the unitary operator for the scalar product of the
following type:

\[
\left( \Phi ,\Psi \right) =\int d\overrightarrow{x}\cdot \Phi ^{\dagger
}\cdot \Psi \mbox{.} 
\]

Therefore:

\[
\widehat{H}=\beta _x\cdot \left( i\cdot \partial _x\right) +\beta _y\cdot
\left( i\cdot \partial _y\right) +\beta _z\cdot \left( i\cdot \partial
_z\right) -i\cdot \widehat{Q}\mbox{.} 
\]

The unitary operator $\widehat{U}\left( \triangle t\right) $ is denoted as
the natural evolution operator if the operator $\widehat{U}\left( \triangle
t\right) $ space is invariant for the complete Poincare group
transformations.

For every natural evolution operator the real number $m$ exists for which
all elements $\Psi $ of this operator space obey to the Klein-Gordon
equation:

\[
\frac{\partial ^2\Psi }{\partial t^2}-\frac{\partial ^2\Psi }{\partial x^2}-%
\frac{\partial ^2\Psi }{\partial y^2}-\frac{\partial ^2\Psi }{\partial z^2}%
=-m^2\cdot \Psi \mbox{.} 
\]

Therefore in this case:

\[
\begin{array}{c}
\widehat{Q}\cdot \widehat{Q}=-m^2\mbox{,} \\ 
\widehat{Q}\cdot \beta _x=-\beta _x\cdot \widehat{Q}\mbox{,} \\ 
\widehat{Q}\cdot \beta _y=-\beta _y\cdot \widehat{Q}\mbox{,} \\ 
\widehat{Q}\cdot \beta _z=-\beta _z\cdot \widehat{Q}\mbox{.}
\end{array}
\]

If

\[
-i\cdot \widehat{Q}=m\cdot \gamma 
\]

then some real number $\alpha $ exists for which:

\[
\gamma =\cos \left( \alpha \right) \cdot \gamma ^0+\sin \left( \alpha
\right) \cdot \beta ^4 
\]

and

\[
\widehat{H}=\beta _x\cdot \left( i\cdot \partial _x\right) +\beta _y\cdot
\left( i\cdot \partial _y\right) +\beta _z\cdot \left( i\cdot \partial
_z\right) +m\cdot \gamma \mbox{.} 
\]

The probability is denoted as the elementary free lepton probability if for
this probability density vector the spinor $\Psi $ exists which is element
of some natural evolution operator space and which obeys to (\ref{pp}).

Therefore the physics elementary particle behavior in the vacuum looks like
to the trackelike probability behavior. That is if the partition with two
slits between the source of the physics particle and the detecting screen
exists in the vacuum then the interference of the probability is observed.
But if this system shall be placed in the Wilson cloud chamber then the
particle shall got the clear trace, marked by the condensate drops, and
whole interference shall vanished. It looks like to the following: the
physics particle exists in the moment, only, in which some event on this
particle is happening. And in other times this particle does not exist and
the probability of some event on this particle exists, only.

Hence, if events on this particle do not happen between the event-birth and
the event-detection then the particle behavior is the probability behavior
between these events, and the interference is visible. But in the Wilson
cloud chamber, where the ionization acts are form the almost continuous
line, the particle has got the clear trace and no the interference. And the
particle moves because such line is not absolutely continuous. Every point
of the ionization act has got the neighboring ionization point, and the
event on this particle is not happen between these points. Therefore, the
physics particle moves because the corresponding probability is propagated
in the space between these points.

\end{document}